# *In-situ* Formation of Superconducting FeTe/layered-MnTe heterostructures

Zhihao He[1], Chen Ma[1], Jiannong Wang[1,2]*, and Iam Keong Sou[1,2]*

[1] *Department of Physics, the Hong Kong University of Science and Technology, Hong Kong 999077, China*

[2] *William Mong Institute of Nano Science and Technology, the Hong Kong University of Science and Technology, Hong Kong 999077, China*

*\* Corresponding authors*

## Abstract

Manganese telluride (MnTe) has garnered strong interest recently for its antiferromagnetic semiconductor properties, which are promising for applications in spintronics, data storage, and quantum computing. In this study, we discovered that the deposition of FeTe at 300°C onto zinc-blende MnTe (ZB-MnTe) via molecular beam epitaxy (MBE) results in a phase transition from ZB-MnTe to a layered MnTe (l-MnTe) phase with van der Waals (vdW) gaps, which is a previously unreported phase of MnTe. The l-MnTe phase was characterized using cross-sectional high-angle annular dark-field (HAADF) imaging, energy-dispersive X-ray spectroscopy (EDS) mapping, and X-ray photoelectron spectroscopy (XPS). The Fe/Te flux ratio during FeTe deposition was found to be critical to the phase transition, an increased Fe/Te flux ratio used for the FeTe growth leads to localized formation of layered $Mn_4Te_3$ (l-$Mn_4Te_3$), while a decreased Fe/Te flux ratio only generates a single monolayer of l-MnTe at the interface and the rest turns into a distorted ZB-MnTe (dZB-MnTe) phase. It was also found that FT-MT heterostructures grown at a lower substrate temperature of 250°C, as the Fe/Te flux ratio decreases, the ZB-MnTe layer was first transformed to dZB-MnTe and then to wurtzite MnTe (WZ-MnTe). The FeTe/l-MnTe heterostructure exhibits high-quality superconducting properties with a three-dimensional nature as demonstrated by its magneto-transport properties and there is evidence that l-MnTe seems to play a key role in inducing the observed superconductivity. Most importantly, this study reports the realization of layered structures of MnTe by an in-situ approach via chemical interactions, which might be further applied to generating unprecedented phases of materials under certain conditions.

## Introduction

Manganese telluride (MnTe) has attracted considerable attention in recent years due to its unique properties as an antiferromagnetic semiconductor, positioning it as a promising material for applications in magnetism [1, 2], spintronics [3-5] and optoelectronics [6, 7], etc. Prior to this study, MnTe is known to exhibit three distinct crystalline phases: nickeline MnTe (NC-MnTe), zinc blende (ZB-MnTe), and wurtzite (WZ-MnTe), each with unique electronic structures and physical properties. NC-MnTe with two interpenetrating hexagonal sublattices is the most stable phase among the three at ambient conditions [8]. It has an indirect bandgap of 1.37-1.51 eV [6] and a high Néel temperature of 307-310K [9-11], which makes it suitable for spintronic devices operating at room temperature [12]. In comparison, ZB-MnTe has a wider indirect bandgap, estimated to be about 3 eV [13, 14], and a lower Néel temperature of about 65 K [15]. ZB-MnTe has been utilized in the fabrication of dilute magnetic semiconductors (DMS) due to its strong antiferromagnetic nearest-neighbor interactions [16]. WZ-MnTe exhibits a direct bandgap of approximately 2.7 eV [8], rendering it of considerable interest for its applications in wide bandgap optoelectronic devices and as a potential back contact layer for CdTe solar cells [17, 18]. The Néel temperature of WZ-MnTe was scarcely documented, only a conference abstract mentioned that it might have a Néel temperature of about 60 K [19]. A previous study provides theoretical prediction that the total energies of ZB-MnTe and WZ-MnTe are 0.25 eV per atom pair and 0.32 eV per atom pair higher than that of NC-MnTe, respectively, which was supported by the x-ray diffraction (XRD) analysis performed on molecular beam epitaxy (MBE)-grown MnTe thin films on $Al_2O_3$ (0001) substrates, showing that all the three phases co-exist with the NC phase being the dominating one [20].

MBE is a sophisticated technique capable of fabricating high-quality single-crystalline thin films and heterostructures. The epitaxial growth of MnTe using MBE has been the subject of extensive research, revealing that substrate selection plays a significant role in determining the resultant phase of MnTe. NC-MnTe has been demonstrated to be grown on α-Al$_2$O$_3$(0001) [21], InP(111)A, and SrF$_2$(111) [5]. For ZB-MnTe, the choice of substrate in the epitaxial growth influences the resulting crystallographic orientation. The orientation of ZB-MnTe(100) is achievable on substrates such as GaAs(100) [22] and ZnTe(100) [23]. ZB-MnTe(111) could be grown on CdTe(111) [14], SrTiO$_3$(001) [24], BaF$_2$(111) [25], and mica [26]. Recently it was demonstrated that WZ-MnTe can be fabricated via magnetron sputtering on a thermally oxidized SiO$_2$ substrate [27].

Despite extensive research has been dedicated to synthesizing these three bulk phases of MnTe, current research has not yet provided conclusive experimental evidence for synthesis of other structural phases of MnTe, especially those with two-dimensional (2D) characteristics. Since the emergence of graphene, 2D materials have attracted scientific attention due to their unique properties, such as strong excitonic effects [28], valleytronics [29], unconventional superconductivity [30], charge-density wave states [31], and complex topological properties [32]. In this context, exploring novel MnTe phases, especially those with layered structures, is of substantial importance due to the potential to uncover novel physical properties and broaden the spectrum of technological applications of this material system. A synthesis approach for creating monolayer 2D MnTe has been documented, involving the reduction of NC-MnTe to just a few atomic layers through a process that integrates melting and liquid phase exfoliation [33]. Despite being reduced in thickness down to monolayer 2D structures, this synthesized MnTe did not exhibit the defining vdW gaps that are characteristic of typical layered 2D materials.

One of our recent works found that the MBE growth of FeTe on Bi$_2$Te$_3$ under specific growth conditions could facilitate the extraction of Te from Bi$_2$Te$_3$, leading to the transformation from Bi$_2$Te$_3$ to Bi$_4$Te$_3$ and Bi$_6$Te due to the strong reactivity between Fe and Te. Additionally, it was found that lowering the FeTe growth temperature could prevent the Te extraction from the Bi$_2$Te$_3$ layer, thereby preserving the Bi$_2$Te$_3$ structure [34]. In the present study, we discovered that a growth process in which the epitaxial growth of FeTe onto a ZB-MnTe layer by MBE induces a transformation from ZB-MnTe into a previously unreported layered MnTe (l-MnTe) phase with vdW gaps. The formation of l-MnTe was verified using a number of structural and chemical analytical techniques. By tuning the substrate temperature and Fe/Te flux ratio for the FeTe growth, transformations from ZB-MnTe to layered Mn$_4$Te$_3$ (l-Mn$_4$Te$_3$), distorted ZB-MnTe (dZB-MnTe) and WZ-MnTe could also be achieved. Furthermore, the resulting FeTe/l-MnTe heterostructure exhibits superconductivity that predominantly reflects a bulk behaviour as evidenced from its magneto-transport properties and the l-MnTe component seems to be the source for the induced superconductivity in this heterostructure system.

## Methods

### Materials Synthesis

All FeTe-MnTe (FT-MT) samples analysed in this study were synthesised using a VG-V80H MBE system equipped with a reflected high-energy electron diffraction (RHEED) facility for in-situ monitoring. Prior to the growth, epi-ready semi-insulating GaAs(111)B substrates were preheated to 580°C to remove passive oxidation from the surface. Subsequently, a ZnSe buffer layer of approximately 100 nm was deposited using a ZnSe compound source. For all FT-MT samples, the MnTe layers were grown at approximately 370°C for 40 minutes using high-purity manganese flakes (99.95%) and tellurium pieces (99.9999%) co-evaporated with cell temperatures of $T_{Mn}$ =740°C and $T_{Te}$ = 280°C. Subsequently FeTe was grown on MnTe via co-evaporation of high-purity iron lumps (99.95 %) and tellurium pieces (99.9999%). The FeTe layers of FT-MT-1, FT-MT-2, and FT-MT-3 were grown at a substrate temperature of 300°C for 60 minutes, using the same Te cell temperature of 280°C but with different Fe cell temperatures of 1175°C, 1180°C, and 1170°C, respectively. For FT-MT-4, FT-MT-5, and FT-MT-6, the FeTe layers were grown at a lower substrate temperature of 250°C for 60 minutes, with a reduced Te cell temperature at 250 ℃ and lower Fe cell temperatures set as 1160°C, 1155°C, and 1150°C, respectively. The growth parameters of the FeTe layers for the FT-MT heterostructures used in this study are summarized in Table I for clarity. Following the growth of FeTe, a ZnSe capping layer of approximately 20 nm was then grown at around 40°C to prevent oxidation of the heterostructures. A pure MnTe sample named MnTe-1 was fabricated using the same growth conditions established for the MnTe layers of the FT-MT heterostructures, however, without the further growth of FeTe.

Table I. Growth parameters of the FeTe layers of the FT-MT samples used in this study

| Sample No. | FeTe | | | |
|---|---|---|---|---|
| | $T_{sub}$ (°C) | $T_{Fe}$ (°C) | $T_{Te}$ (°C) | Growth time (min) |
| FT-MT-1 | 300 | 1175 | 280 | 60 |
| FT-MT-2 | 300 | 1180 | 280 | 60 |
| FT-MT-3 | 300 | 1170 | 280 | 60 |
| FT-MT-4 | 250 | 1160 | 250 | 60 |
| FT-MT-5 | 250 | 1155 | 250 | 60 |
| FT-MT-6 | 250 | 1150 | 250 | 60 |

## Materials Characterization

The crystalline properties of the samples used in this study were monitored during the growth using the reflection high-energy electron diffraction (RHEED) facility. High-resolution X-ray diffraction (HRXRD) measurements were performed using a PANalytical Multipurpose X-ray Diffractometer equipped with Cu Kα1 X-rays (wavelength of 1.54056 Å). To reveal the lattice structure of the FT-MT samples and the MnTe sample, cross-sectional high-resolution spherical-aberration-corrected scanning transmission electron microscopy (STEM) imaging was conducted using a JEM-ARM200F transmission electron microscope operating at 200 keV. The STEM system was equipped with a probe corrector and a HAADF detector, which enabled detailed analysis of the crystal structure. STEM samples were prepared by the focused ion-beam (FIB) lift-out technique using a Zeiss DualBeam Cross Beam uSTE750 at 30 kV. The energy dispersive spectrometer (EDS) was performed for elemental mapping using a built-in Bruker EDS system consisting of four silicon drift detectors in the Themis microscope. The measurements of the X-ray photoelectron spectroscopy (XPS) spectra were conducted using a Kratos-Axis Ultra DLD XPS *ex situ*. The ion sputtering performed during the XPS depth profiling was handled using Ar ions with 4 kV with 3 mm × 3 mm raster and 140 µA extractor current.

The FT-MT samples were cut into a long strip with dimensions of approximately 2 mm × 4 mm using a diamond scriber, and the electrical contacts were created by bonding aluminium wires onto the sample surface by a wire bonder. The magneto-transport properties of these samples were then measured in a Quantum Design physical property measurement system using the 4-prob technique from 300 K to 2K.

## Results and discussion

Figure 1 displays the RHEED patterns obtained during the growth of FT-MT-1 and the HRXRD profiles of MnTe-1 and FT-MT-1. As shown in Figure 1(a), the RHEED patterns of the MnTe layer of FT-MT-1 displayed a narrower set of streaks when the incident electron beam was along the sample rotational angle at $\varphi=0°$ and a wider set of streaks at $\varphi=30°$. The spacing ratio for the two sets of streaks (marked in red and blue) is about $1:\sqrt{3}$. The streaky patterns exhibited a 60° rotational symmetry, indicating that the in-plane atomic structure of the as-grown MnTe layer has a six-fold symmetry. Figure 1(b) shows the RHEED patterns of the as-grown FeTe layer of FT-MT-1 at rotational angles $\varphi=0°$, $\varphi=15°$, and $\varphi=30°$. These diffraction patterns display a 30° rotational symmetry. The narrower set of streaks and the wider set of streaks occur at 0° and 15° respectively, with a spacing ratio of about $1:\sqrt{2}$ (marked in red and green in Figure 1(b)). These RHEED observations are similar to those observed in the FeTe/Bi-Te heterostructures previously reported by our group [34] and in FeTe:Se/$Bi_2Te_3$ heterostructures by others [35]. These RHEED observations were attributed to the epitaxial growth of three tetragonal lattices in as-grown FeTe layer with the twisted angles of 0°, 60°, and 120°, respectively. Additionally, a lattice spacing ratio of $\sqrt{3}:2 \approx 7:8$ between the FeTe layer and Bi-Te layer was demonstrated as the reason why FeTe and Bi-Te exhibit different lattice symmetries but can undergo epitaxial growth. In alignment with these findings, the in-plane lattice constants of FeTe and ZB-MnTe during the onset of the FeTe growth of the FeTe/MnTe heterostructures, as calculated from the spacings of their RHEED's streaks, were approximately 3.8 Å and 4.38 Å, which also enjoy a ratio of about $\sqrt{3}:2$, thus explaining that why the FeTe layers in FT-MT samples also display a 30° rotational symmetry. As mentioned earlier, MnTe has three common phases: nickeline (NC), zinc blende (ZB), and wurtzite (WZ). In NC-MnTe(001), ZB-MnTe(111), and WZ-MnTe(001), they all exhibit in-plane hexagonal lattice structures. Aiming to determine the phase of the as-grown MnTe layers in the FT-MT heterostructures, we conducted HRXRD profiling on a sample named MnTe-1 that was grown under the same growth conditions as those used in FT-MT heterostructures but without proceeding the growth of FeTe on it. The upper part of Figure 1(c) shows the HRXRD results of MnTe-1, displaying only the characteristic diffraction peaks of ZB-MnTe (111) and ZB-MnTe (222) in addition to the peaks from the substrate and ZnSe buffer layer. These measured values of 2theta for the characteristic peaks from MnTe-1 are not only very close to those recorded in the crystalline database [36] but also align well with other reports [24] [37], indicating that the phase of the MnTe layers grown on ZnSe/GaAs(111) substrates under our growth conditions is ZB-MnTe(111). The bottom part of Figure 1(c) displays the HRXRD profile of FT-MT-1, in which one can see that in addition to the peaks from the substrate, ZnSe buffer layer and MnTe layer, it displays characteristic diffraction peaks of FeTe (00l), FeTe (002), FeTe (003), and FeTe (004), confirming the epitaxial growth of tetragonal FeTe on MnTe in this heterostructure.

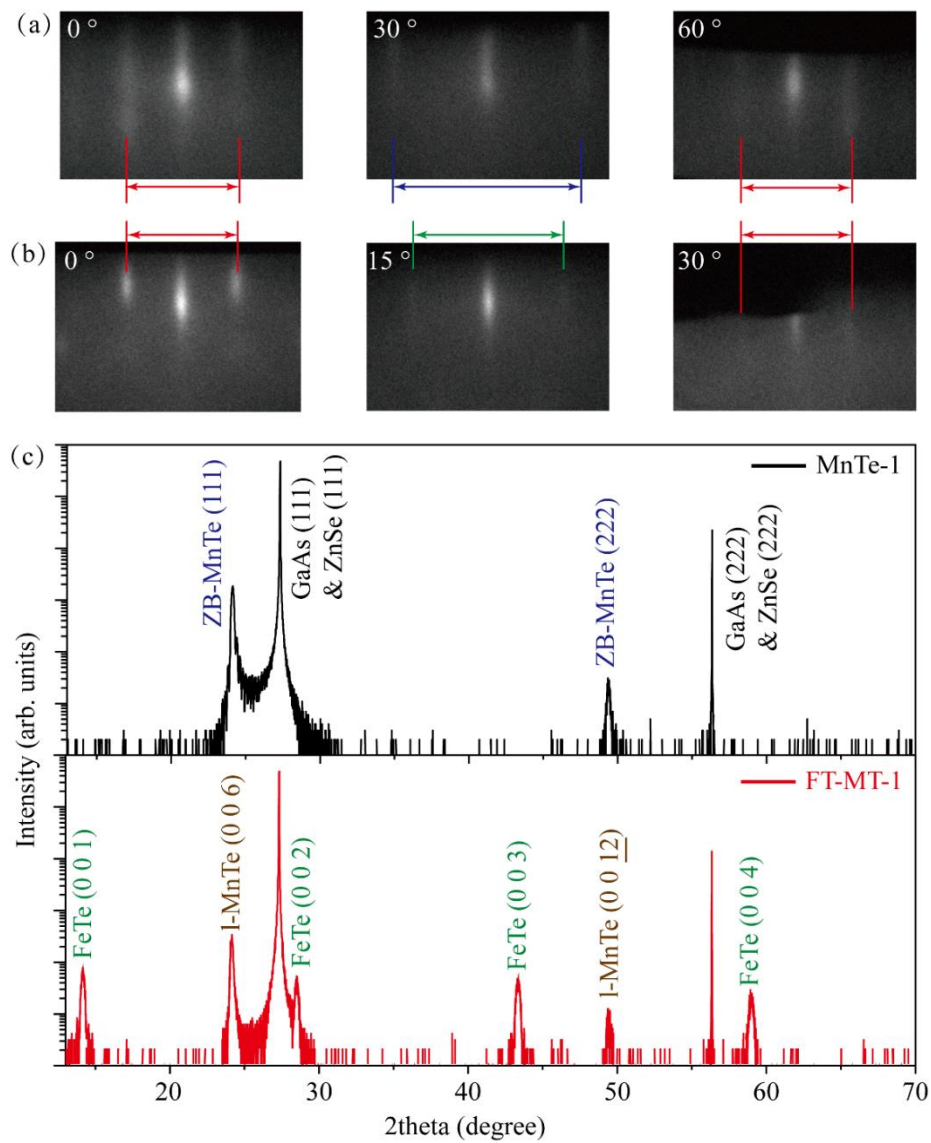

Figure 1. RHEED patterns of (a) MnTe layer and (b) as-grown FeTe layer in FT-MT-1. (c) High-resolution X-ray diffraction (HRXRD) profiles of MnTe-1 (top) and FT-MT-1 (bottom) from 13° to 70°.

To obtain the detailed information on the structural properties of FT-MT-1, cross-sectional scanning transmission electron microscopy (STEM) imaging with a HAADF detector was performed. Interestingly, HAADF images of the FT-MT-1 heterostructure reveal that the MnTe layer of FT-MT-1 displays an intriguing new structural phase. As shown in Figure 2(a) (b), the HAADF images captured at the MnTe layer region and the interface between FeTe and MnTe indicate that the new-phase MnTe has a layered lattice structure, which we hereafter refer to as layered MnTe (l-MnTe). The l-MnTe is characterised by a uniform atomic arrangement with distinct vdW gaps between the neighbouring layers. Figure 2(c) presents a magnified view of the marked area in Figure 2(a), where a single layer is observed to contain a double layer of Mn atoms (the smaller spheres) interposed between two layers of Te atoms (the larger and brighter spheres). Building upon this visual evidence, the atomic structure of l-MnTe is schematically depicted in Figure 3(d), showing that each MnTe layer consists of two staggered honeycomb MnTe sublayers, which are covalently bonded via Mn–Te linkages. This structure is similar to that of the monolayer MnSe, as reported by Kehan Liu et al. [38] and Markus Aapro et al. [39] The HAADF image reveals an interlayer spacing $d_c \approx 7.32$ Å, and the spacing between the two nearest neighbouring Te atoms along the <120> direction is $d_{<120>} \approx 3.75$ Å. From this $d_{<120>}$ value, the in-plane lattice constants are deduced to be about a=b=4.33 Å. As indicated in Figure 2(c), the unit cell of l-MnTe might comprise three monolayers, with a lattice constant along the c-axis measured to be 21.96 Å. By using the established c-axis lattice constant from the HAADF images, and via Brag's law, we deduced that the diffraction peak position of l-MnTe(006) should approximately have a 2theta value of about 24.29°, which is very close to that of the ZB-MnTe(111) peak (at 24.202°, extracted from the crystalline database [36]). Such a close similarity in diffraction peak positions provides an explanation why the HRXRD profile of FT-MT-1 displays diffraction peaks of l-MnTe similar to those of ZB-MnTe(111). These diffraction peaks observed in the HRXRD profile of FT-MT-1 correspond to the diffraction peaks of l-MnTe with indices of (0 0 <u>6</u>) and (0 0 <u>12</u>) as marked in Figure 1(c). It is worth mentioning that a careful inspection of the HAADF images shown in Figure 2(a) and (b) reveals that the l-MnTe lattice in FT-MT-1 contains two different domains along the c-axis with a difference in their in-plane orientation of 180 degrees. In these images, we have marked down the boundaries of these two domains by blue arrows. Such a domain structure along the c-axis may be simply considered as a change of the stacking order triggered by the instability of the l-MnTe lattice during its formation.

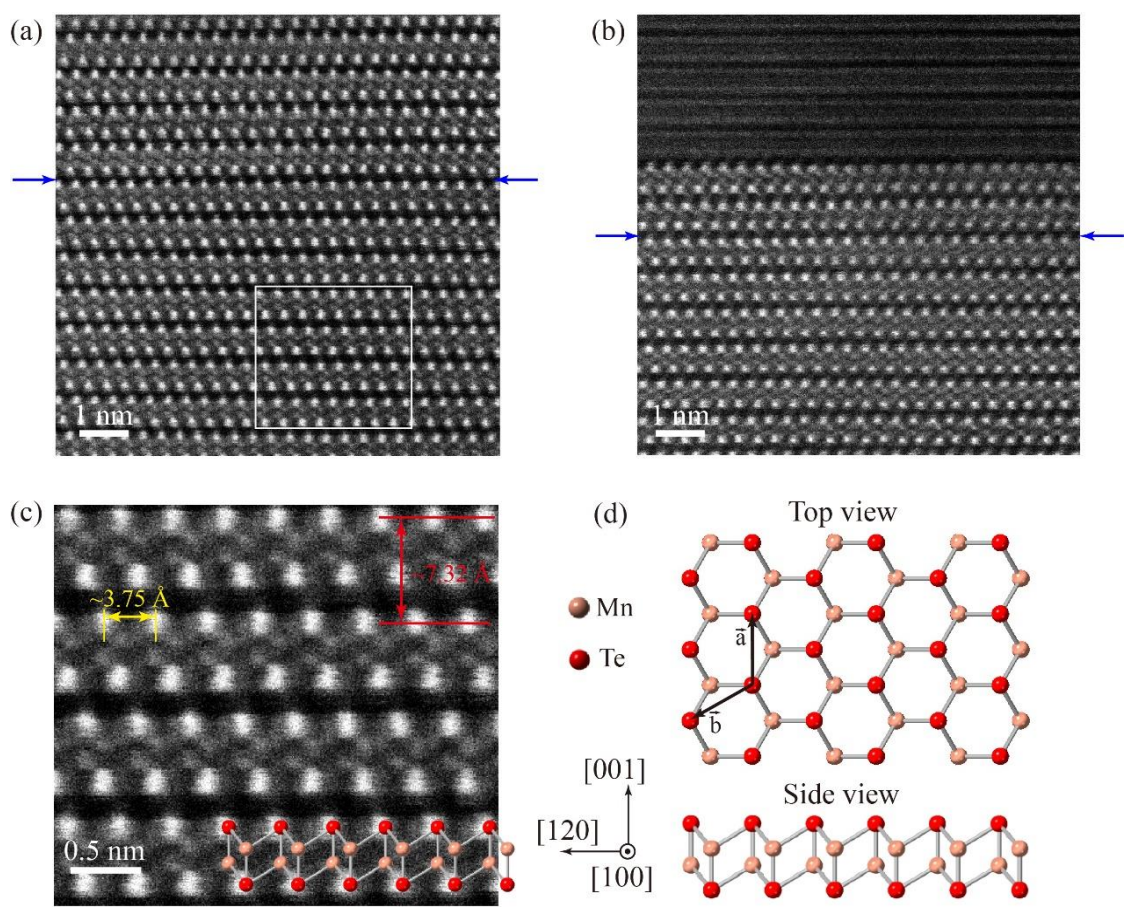

Figure 2. Structural analysis of FT-MT-1. The cross-sectional HAADF images captured at (a) the MnTe layer region and (b) the interface between FeTe and MnTe. (c) An enlarged view of the area marked in (a) . (d) Schematic drawings of the atomic structgure of l-MnTe.

To investigate the elemental compositions of the layers within FT-MT-1, energy-dispersive X-ray (EDX) spectroscopy mapping was conducted on a cross-sectional scanning transmission electron microscopy (STEM) image of this sample. As illustrated in Figure 3(a), the thicknesses of the FeTe layer and the MnTe layer are 32nm and 60nm, respectively. The EDX mapping results reveal that Fe and Mn are primarily localized within their corresponding FeTe and MnTe layers, respectively, offering evidence that interlayer diffusion is insignificant. However, this cross-sectional STEM image together with the EDX mapping encompass these two distinct layers together within the sampling region, which brings difficulty in determining the elemental stoichiometry of the MnTe layer. To obtain more accurate compositional data, X-ray photoelectron spectroscopy (XPS) analysis was performed on the MnTe layer of another piecce of FT-MT-1 after removing the ZnSe capping layer and the FeTe layer via the sputtering with an Ar$^+$ ion source. During the sputtering process, the XPS signals of Zn, Se, Fe, Mn and Te were monitored so as to confirm the complete removal of the ZnSe capping layer and the FeTe layer prior to taking the XPS profiles for the remaining MnTe layer. Figures 3(b) and (c) display the obtained XPS spectra of Mn 2p and Te 3d core-levels, respectively. These XPS spectra include the experimental data, the fitted Shirley background and the resolved component peaks. As depicted in Figure 3(b), the Mn $2p_{3/2}$ core-level spectrum can be deconvoluted into four distinct peaks. Peak 1 at 639.7 eV is assigned to Mn in a low charge state, $Mn^{q+}$ ($0 < q < 1$) [40]. This peak is potentially associated with the formation of manganese subtelluride nanoclusters as a consequence of Ar$^+$ ion sputtering [40-42]. Peak 2 at 640.8 eV is contributed by the $Mn^{2+}$ state within Mn–Te bonds [41, 42], which is expected to be common for the various phases of MnTe including the layered phase, as they all involve similar Mn-Te covalent bonds. Peak 3 and Peak 4 are the associated satellite peaks of Peak 1 and Peak 2, respectively. [41-43] These satellite peaks arise from charge transfer interactions between the ligand's outer electron shell and the unfilled 3d shell of Mn, occurring concurrently with the generation of a core hole during the photoemission process [40, 44]. In Figure 3(c), the peaks at 583 eV and 572.6 eV are assigned to the Te $3d_{3/2}$ and $3d_{5/2}$ orbitals with a splitting of 10.4 eV. These two peaks are associated with the $Te^{2-}$ state in Mn–Te bonds [41, 42]. Based on the peak areas and the relative sensitivity factors from the XPS data, the Mn and Te chemical composition are estimated to be about 53.29% and 46.71% respectively, yielding a Mn:Te atomic ratio to be about 1:1, confirming that the l-MnTe layer in FT-MT-1 is close to stoichiometric, being consistent with the structure revealed by the STEM images shown in Figure 2.

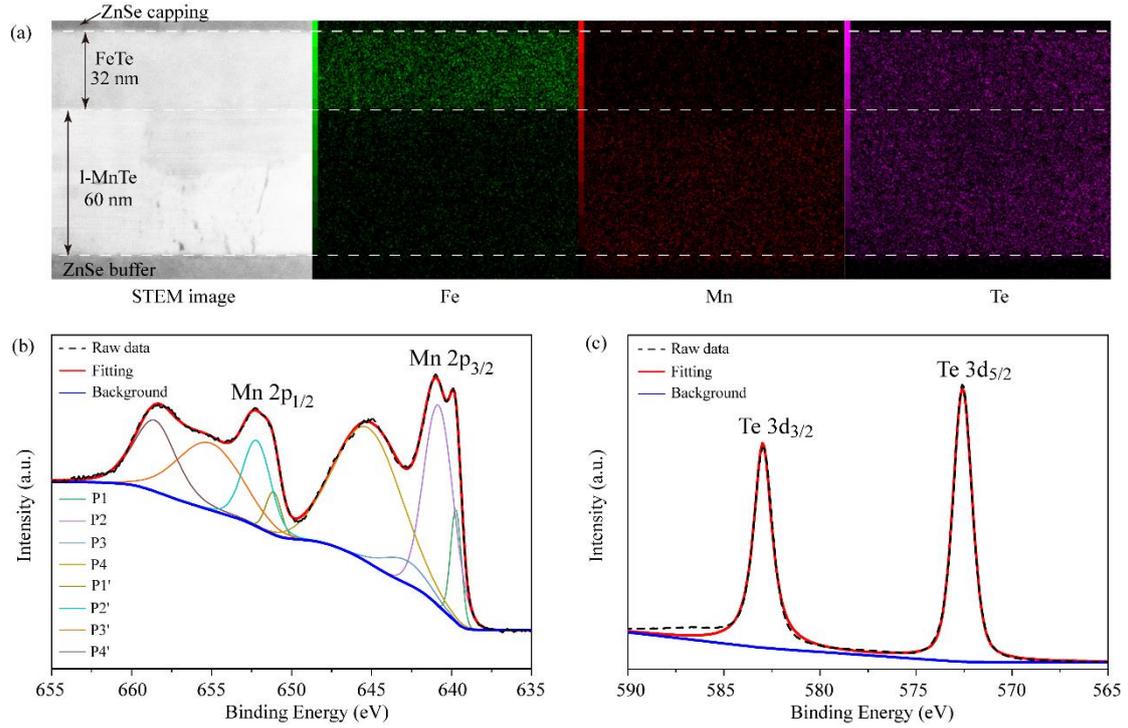

Figure 3. Elemental analysis of FT-MT-1. (a) The cross-sectional STEM image of FT-MT-1 and corresponding EDS mapping with uniform distribution of Fe, Mn, and Te atoms, marked as the regions between the white dash lines. (b) and (c) display the X-ray photoelectron spectroscopy (XPS) spectra of the Mn 2p, and Te 3d core-levels, respectively, of FT-MT-1 performed after the removal of the ZnSe capping and the FeTe layer by sputtering.

Here comes an interesting question regarding if the l-MnTe layer of the FT-MT-1 sample was formed before the growth of its FeTe layer or it was formed in-situ during the growth of the FeTe layer. As mentioned earlier, the l-MnTe(0 0 6) and l-MnTe(0 0 12) peaks display 2theta values very similar to those of ZB-MnTe(111) corresponding peaks in HRXRD profiles, it is thus impossible to determine the layer phase of the MnTe-1 sample via just the HRXRD profiling. We then conducted the cross-sectional HAADF imaging on MnTe-1. Figure 4(a) displays the resulting HAADF image of MnTe-1 viewed from the <1$\bar{1}$0> direction, where brighter and larger spheres represent Te atoms, while small spheres represent Mn atoms, respectively. Figure 4(b) shows an enlarged view of the area marked in Figure 4(a), offering a more detailed examination to ascertain the phase of MnTe. Accompanied by the atomic structure schematics of ZB-MnTe(111) shown in the top half of Figure 4(b), it can be confirmed that the MnTe phase in MnTe-1 is ZB-MnTe. This suggests that the MnTe layer undergoes a phase transformation from ZB-MnTe(111) to l-MnTe(001) during the FeTe growth of FT-MT-1. As demonstrated in our previous work, FeTe grown on $Bi_2Te_3$ induces the phase transformation of $Bi_2Te_3$ due to the strong reactivity between Fe and Te atoms [34]. This cause may be extended to explain the phase transition observed in this study for the FeTe/l-MnTe heterostructure, however, further studies are required to reveal its detailed mechanism.

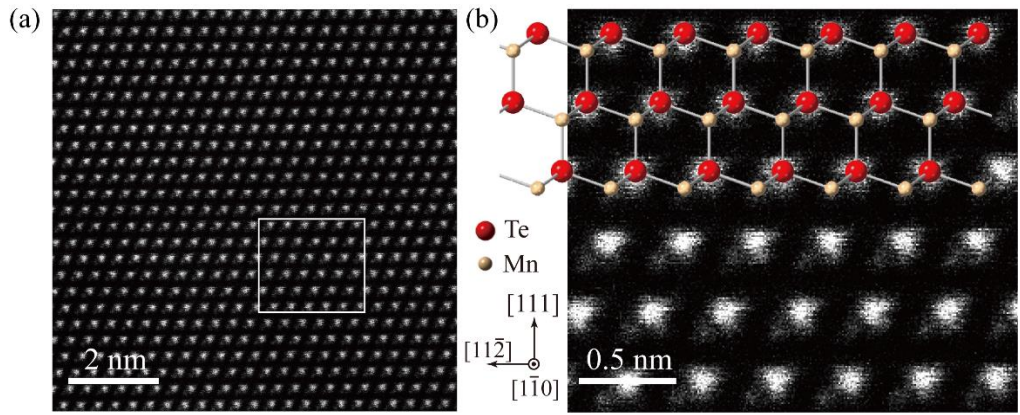

Figure 4.(a) The cross-sectional HAADF image of MnTe-1 viewed from <1$\bar{1}$0>. (b) An enlarged view of the area marked in (a) and atomic structure schematics of ZB-MnTe(111).

Investigation of the dependence of the Fe/Te flux ratio on the phase transformation of the underlying ZB-MnTe layer during the growth of FeTe was also carried out. We have fabricated two more samples: FT-MT-2 and FT-MT-3, which were grown using different Fe cell temperatures of 1180°C and 1170°C respectively for the growth of the top FeTe layer, while keeping other conditions the same as those used in fabricating FT-MT-1. Here it should be restated that the Fe cell temperature used for the growth of the FeTe layer of FT-MT-1 is 1175°C.

Figures 5(a) and (b) present the cross-sectional HAADF images of FT-MT-2 at the FeTe/MnTe interface and within the MnTe layer, respectively. As shown in these figures, the MnTe layer in FT-MT-2 also exhibits a layered structure. Being different from the l-MnTe observed in FT-MT-1, even though FT-MT-2 shows dominating l-MnTe lattice, however, within its HAADF images three regions (confined by the 3 pairs of red lines in Figure 5(a) and (b)) were found to consist of a layered Mn-Te compound distinct from l-MnTe, characterized by the presence of three Te atomic layers in a single unit. A zoomed-in view of the area marked in Figure 5(b) is presented in Figure 5(c) to offer a more detailed examination of the fine structure for this distinct Mn-Te phase. As shown in Figure 5 (c), this layered structure comprises three Te atomic layers (the larger, brighter spheres), with a double layer of Mn atoms (the smaller spheres) interposed between two adjacent Te layers, resulting in a layered $Mn_4Te_3$ (l-$Mn_4Te_3$) compound. Building upon this visual evidence, the atomic structure of l-$Mn_4Te_3$ was schematically constructed, which is shown in Figure 5(d). The HAADF image shown in Figure 5(c) also reveals that an interlayer spacing of approximately $d_c$ =12 Å, and the spacing between the two nearest neighbouring Te atoms along the <120> direction is $d_{<120>}$ = 3.71 Å. From this $d_{<120>}$ value, the in-plane lattice constants are deduced to be about a=b=4.3 Å.

The above structural analysis for FT-MT-2 reveals that an increased Fe cell temperature, that is a higher Fe flux equivalently, may induce the transformation of some l-MnTe into l-$Mn_4Te_3$, which may be attributed to a stronger Fe interaction that could extract a Te atomic layer within two l-MnTe units to form a single unit of l-$Mn_4Te_3$. It should be pointed out that in Figure 5(b), one could also detect the boundaries of the two domains of l-MnTe as marked by the blue arrows in these HAADF images.

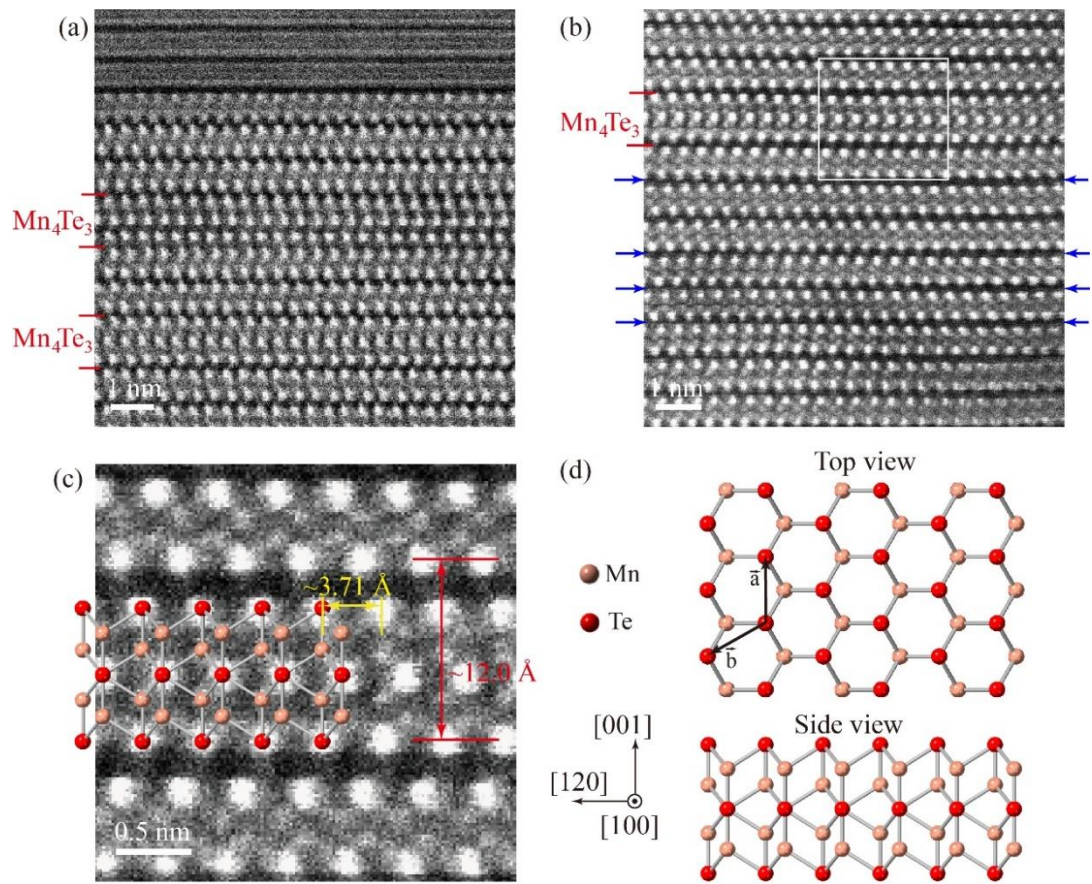

Figure 5. Structural analysis of FT-MT-2. The cross-sectional HAADF images captured at (a) the interface and (b) the MnTe layer region. (c) An enlarged view of the area marked in (b). (d) Schematic drawings of the atomic structure of l-$Mn_4Te_3$.

Figure 6 displays the HAADF images of FT-MT-3, in which the top FeTe layer was grown using an Fe cell temperature lower than that used for FT-MT-1. As shown in Figure 6(a), it seems that only a single l-MnTe layer was formed at the interface of the heterostructure. Figure 6(b) shows a magnified interface region and indeed it confirms this observation, showing a single l-MnTe at the interface with two vdW gaps located at its top and bottom. In both Figure 6(a) and (b), it can be seen that the remaining lattice of the MnTe layer of FT-MT-3 maintains the ZB-MnTe phase, however, a zigzag deformation of its lattice is clearly evident. We believe that the lower Fe flux used for FT-MT-3 as compared with that used for FT-MT-1 may just provide an interaction strength that is capable of triggering a transition from a ZB-MnTe lattice to a zigzag deformation of the lattice, resulting in the formation of the dZB-MnTe phase, which may be a pre-phase of the l-MnTe phase. However, this interaction is not strong enough to turn the ZB-MnTe lattice into the l-MnTe phase except the top layer next to the interface of the heterostructure.

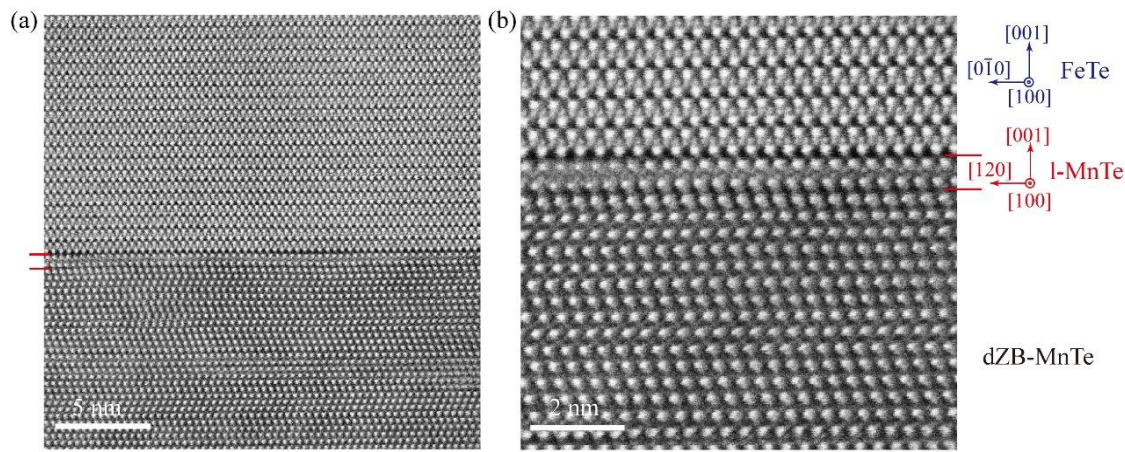

Figure 6. Structural analysis of FT-MT-3. (a) The cross-sectional HAADF image captured at the interface. (b) A high-magnification HAADF image captured at the interface.

Based on these findings in our previous work that lowering the growth temperature to a certain value for the growth of FeTe on $Bi_2Te_3$ could prevent the Te extraction from the $Bi_2Te_3$ layer, thereby preserving the $Bi_2Te_3$ structure [34], we believe that reducing the FeTe growth temperature in fabricating FT-MT heterostructures might also lessen the transformation of the bottom ZB-MnTe. Thus, we fabricated three more samples named FT-MT-4, FT-MT-5, and FT-MT-6, in which, their FeTe layers were grown at 250°C, with Te source temperature at 250°C, while their Fe source temperatures were set at 1160°C, 1155°C, and 1150°C, respectively.

Figure 7(a-c) display the HAADF images taken near the interface regions between the FeTe layer and MnTe layer, while Figure 7(d-f) show the magnified images taken within the MnTe layers, for FT-MT-4, FT-MT-5, and FT-MT-6, respectively. As expected, all the MnTe layers in these three heterostructures do not contain any layered components of MnTe. However, it is interesting to see that the MnTe layer of FT-MT-4 shows a distorted zincblende (dZB) phase while both the MnTe layers of FT-MT-5 and FT-MT-6 display a wurtzite phase. In Figure 7(g), a magnified portion of the image in Figure 7(f) as marked by a square is displayed together with its atomic configuration. These results indicate that among these three heterostructures with the FeTe layer grown at 250°C, when the Fe/Te flux ratio is lower to a certain value, the original ZB-MnTe will be transformed into a wurtzite phase.

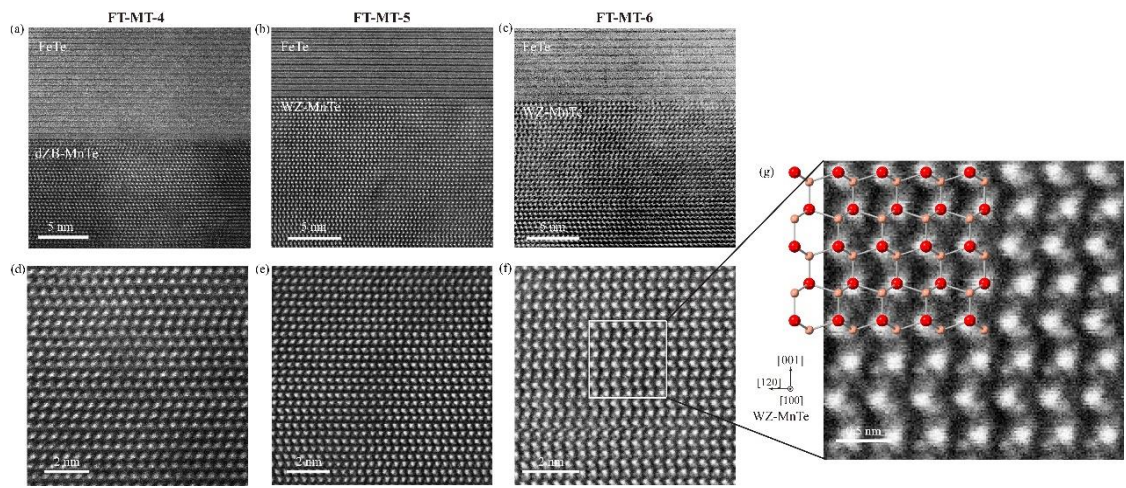

Figure 7. (a-c) The cross-sectional HAADF images captured near the interface for FT-MT-4, FT-MT-5, and FT-MT-6, repsctively; (d-f) Corresponding high-magnification HAADF images captured within the MT layers. (g) An enlarged view of the area marked by a white square in (f) together with the schematic drawings of the atomic structure of WZ-MnTe.

In this paragraph, we would like to address the intrinsic cause why ZB-MnTe could be easily transformed to different phases via the growth of an FeTe layer on it. The TEM specimen used for obtaining the HAADF images of the pure MnTe thin film (MnTe-1) as shown in Figure 4 was stored in a plastic bag filled with high-purity nitrogen gas right after the first imaging study. A second imaging study on the same specimen was performed 39 days later and Figure 8 displays the resulting cross-sectional HAADF images. Fig. 8(a) displays a typical region of this specimen, which shows that most of the original ZB phase has been transformed to dZB phase. More interestingly, as shown in Figiure 8(b), a few layers of l-MnTe and l-Mn$_4$Te$_3$ appear in a rare region of this specimen, as can be seen in the area confined by the red-dash rectangle marked in this image. These observations indicate that ZB-MnTe is not a stable phase and itself can go through a transformation into the dZB-MnTe phase or even l-MnTe and l-Mn$_4$Te$_3$ over a prolonged period of time while the growth of FeTe on top of ZB-MnTe will trigger the phase transformations observed in this study to take place much more quickly and more uniformly.

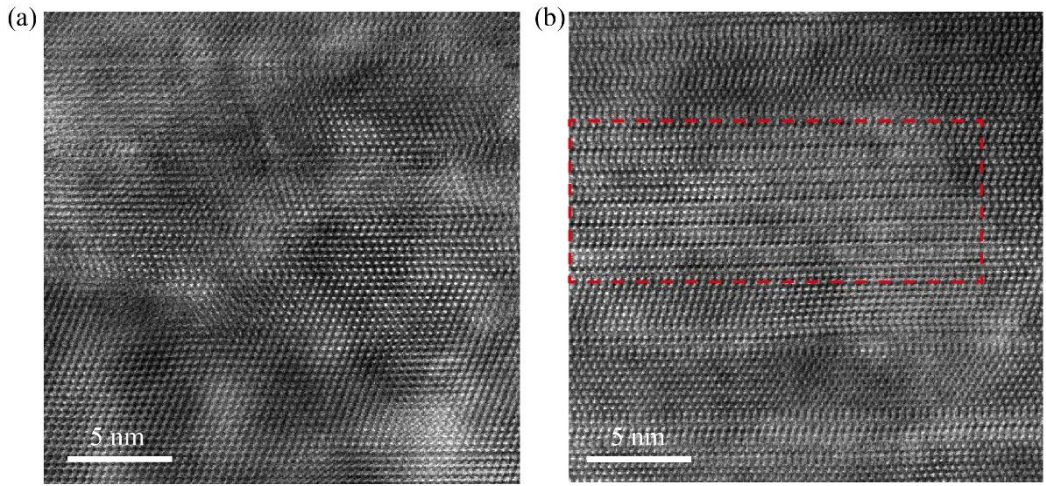

Figure 8. The cross-sectional HAADF images of MnTe-1 taken from the same TEM speciment 39 days after the HAADF images shown in Figure 4 were captured: (a) a typical region showing dZB-MnTe structure; (b) a rare region showing a few layers of l-MnTe and l-Mn$_4$Te$_3$ inside the dZB-MnTe structure.

The magneto transport characterization performed for the FT-MT samples was initiated with the FT-MT-1 heterostructure of which the bottom layer is a l-MnTe thin film. Figure 9(a) shows its temperature dependent resistance from 300 K to 2K. As can be seen, in addition to a fall at around 58K that might be the antiferromagnetic phase transition of l-MnTe, FT-MT-1 displays superconductivity with the resistance dropping at an onset temperature ($T_{onset}$) of around 12 K and reaching a zero-resistance state at about 9 K. Figure 9(b) displays the temperature-dependent resistances of FT-MT-1 under both perpendicular and parallel magnetic fields, in which the observed diminishment of superconductivity with increasing magnetic field strength serves as the confirmation of its superconductivity. Additionally, an anisotropy regarding the direction of the applied magnetic field was observed. To reveal the dimensional characteristics of the superconductivity of FT-MT-1, we studied the critical temperature dependence of the upper critical magnetic field, in which the values of critical temperature ($T_c$) were extracted using two different approaches. The first one involved identifying $T_c$ by finding the intersection of linear extrapolations from the normal-state and superconducting transition, as depicted in the inset of Figure 9(c). The second one involved identifying the point where the resistance dropped to 90% of the normal-state value within the superconducting transition. The resulting plots using the two approaches are shown in Figure 9(c) and Figure 9(d), respectively. In contrast to our previous work on $Bi_2Te_3$/FeTe [45] and $Sb_2Te_3$/$Fe_{1+y}$Te [46] heterostructures where we observed two-dimensional (2D) superconductivities characterized by a linear dependence of $H_\perp(T)$ versus the critical temperature, FT-MT-1 heterostructure shows non-linear dependence for both $H_\perp(T)$ and $H_{//}(T)$ versus critical temperature for both approaches. Thus, the superconductivity observed in the FT-MT-1 might be predominantly three-dimensional (3D) rather than 2D, which indeed deserved further investigation.

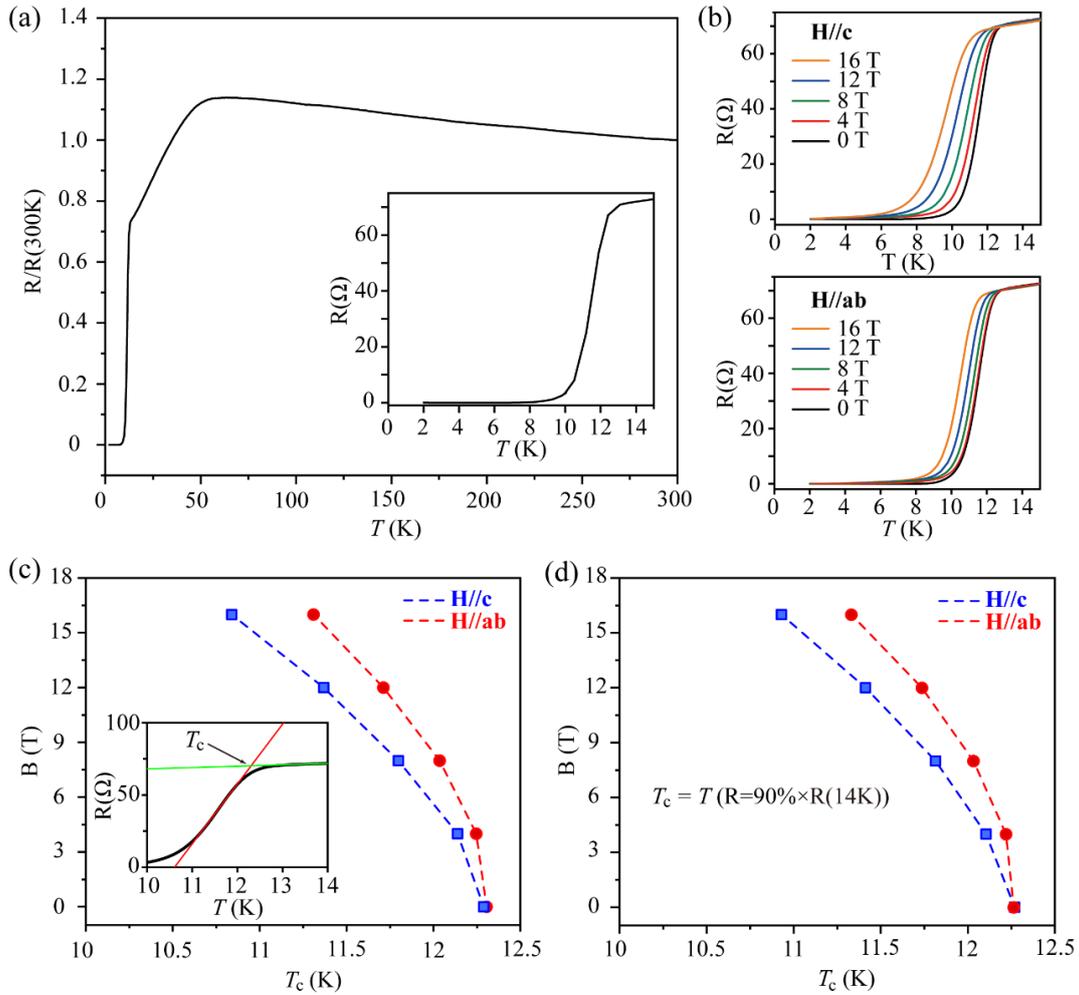

Figure 9. (a) Temperature-dependent resistance of FT-MT-1 from 300K to 2K. (Inset) Temperature-dependent resistance of FT-MT-1 from 15K to 2K; (b) Resistance of FT-MT-1 as a function of temperature under out-of-plane (top) and (bottom) in-plane magnetic field up to 16 T. Temperature dependence of the upper critical fields, where critical temperatures ($T_c$) were determined by (c) the intersection of linear extrapolations from normal-state and superconducting transition (shown in inset), and (d) by the temperature at which the resistance drops to 90% of the normal-state.

Temperature-dependent resistance measurements were also performed for other FT-MT samples. Figure 10 displays the resistance versus temperature (R-T) curves of all FT-MT samples from 300K to 2K. For the three FT-MT samples grown at 300°C, as mentioned earlier, Sample FT-MT-1, displays a superconductivity with resistance reaching zero. The R-T curve for Sample FT-MT-2, of which its FeTe layer was grown with a higher Fe/Te flux ratio than that of Sample FT-MT-1, exhibits a superconducting transition near 10 K but it does not reach a zero-resistance state even at temperature down to 2K. This is consistent with our previous study on the $Sb_2Te_3/Fe_{1+y}Te$ heterostructures that excess Fe in the FeTe layer can hamper the superconductivity quality. The R-T curve of Sample FT-MT-3 does not display superconductivity, instead a rising trend of resistance is observed at its low temperature data. We suspect that the single l-MnTe layer at the interface is not thick enough to induce a superconductivity above our measurement limit of 2K, assuming the observed superconductivity in FT-MT-1 and FT-MT-2 is due to the formation of the l-MnTe structure. For the samples with the FeTe layer grown at a lower substrate temperature of 250 C (FT-MT-4, FT-MT-5 and FT-MT-6), none of their R-T curves display superconductivity.

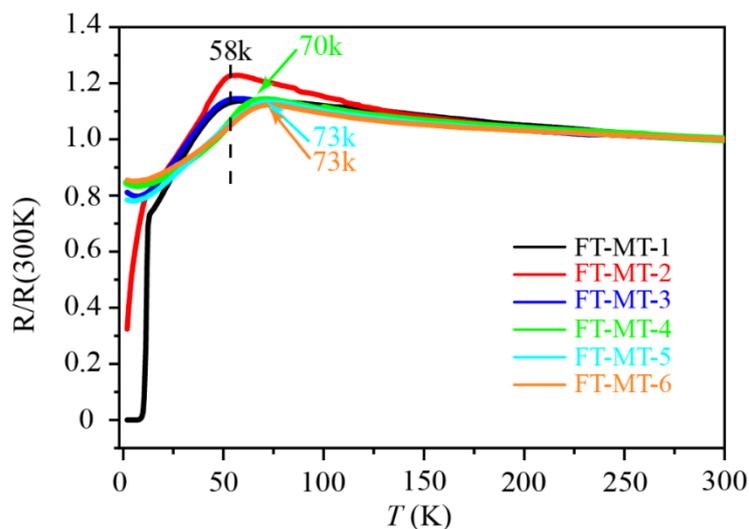

Figure 10. Temperature-dependent resistances of all FT-MT samples used in this study from 300 K to 2 K.

Notably, a fall of resistance occurs around 58K (indicated by the black dashed line) in the R-T curves of FT-MT-1, FT-MT-2 and FT-MT-3 (all of them have l-MnTe in their bottom MnTe layers), whereas it occurs at 70 K for FT-MT-4 and about the same at 73K for FT-MT-5 and FT-MT-6. It is worth mentioning that our MBE-grown FeTe thin films usually display an antiferromagnetic (AFM) transition around 48K, while for $Bi_2Te_3$/FeTe and $Sb_2Te_3$/FeTe the corresponding fall of resistance may be shifted to slightly higher temperatures [45-46] because the conductance of both $Bi_2Te_3$ and $Sb_2Te_3$ will dominate that of FeTe at low temperatures. Thus, it is likely that all the observed falls of resistance in the R-T curves of the FT-MT samples described above could be assigned to the AFM transition of the bottom MnTe layers with the order of 58 K, 70K and 73K for the l-MnTe, dZB-MnTe and WZ-MnTe, respectively, based on the structural analysis via their HAADF images addressed earlier. However, these assignments cannot be concluded until one could detach the various MnTe layers of these heterostructures without changing their properties and then study their R-T characteristics individually, which is definitely a very challenging task and is being explored in our lab.

Here, we would like to further address the 3D nature of the superconductivity induced in Sample FT-MT-1 and FT-MT-2 or even in Sample FT-MT-3 as it may also be superconducting but at a temperature lower than 2K as all of these three samples contain l-MnTe though Sample FT-MT-3 only has a single l-MnTe at the interface of the heterostructure. For the induction of the 2D superconductivity observed in $Bi_2Te_3$/FeTe and $Sb_2Te_3$/FeTe, our previous studies suggest that the itinerant electrons of the topological interface states of the involved topological insulators in these heterostructures play an important role in supressing the AFM state of FeTe and promote superconductivity pairing [45, 46]. In a recent study by Hemian et al., a similar mechanism with an additional Ruderman-Kittel-Kasuya-Yosida-type ferromagnetic coupling mediated by the topological surface states was also proposed [47]. Since the layered-MnTe phase is unlikely to be a topological insulator, here we suggest that it may have an unusual magnetic property that could offer a non-local or long-distance effect that may suppress the AFM state of the entire FeTe layer so that the induced superconductivty in the FeTe/l-MnTe heterostructures thus behaves like a bulk superconductor, however, further experimental and theoretical studies are required to test this hypothesis.

In conclusion, this study demonstrated that epitaxial growth of FeTe at 300 °C on ZB-MnTe via MBE under certain growth conditions could induce a transformation from ZB-MnTe to l-MnTe, resulting in the formation of an FeTe/l-MnTe heterostructure system. The combined analyses from cross-sectional HAADF imaging, EDS mapping and XPS spectra reveal that the l-MnTe phase has a layered structure with in-plane hexagonal symmetry and exhibits a Mn:Te stoichiometric ratio close to 1:1. The Fe/Te flux ratio used during the FeTe growth was found to be crucial to the phase transformation occurring in the MnTe layer, a higher ratio will lead to localized transformation into l-$Mn_4Te_3$ among the dominating l-MnTe lattice, while lower ratio minimizes the conversion from ZB-MnTe to l-MnTe. Furthermore, by lowering the growth temperature of the top FeTe layer to 250 °C, it was found that the original ZB-MnTe layer will go through transformation to dZB-MnTe and then WZ-MnTe as the Fe/Te flux ratio decreases. The magneto-transport properties of the as-grown FeTe/l-MnTe heterostructure reveal a superconducting transition around 12 K, which displays a three-dimensional nature as revealed by its magneto-transport properties. The results obtained from the R-T measurements of these heterostructures seem to indicate that the l-MnTe component is critical for inducing the observed superconductivity. Most importantly, the current study not only demonstrates the experimental realization of previously undocumented layered phases of MnTe but also provides a novel synthesis pathway for realizing unprecedented phases of materials via in-situ chemical interactions.

## Acknowledgements

Zhihao He and Chen Ma contributed equally to this work. We gratefully acknowledge the use of the facilities in the Materials Characterization and Preparation Facility (MCPF) at the Hong Kong University of Science and Technology. The work described here was substantially supported by funding from the Research Grants Council of the Hong Kong Special Administrative Region, China, under Grant Numbers 16308020 and C6025-19G.